# Use of a porous membrane for gas bubble removal in microfluidic channels: physical mechanisms and design criteria

Jie Xu, Regis Vaillant and Daniel Attinger*

*Laboratory for Microscale Transport Phenomena, Department of Mechanical Engineering, Columbia University, New York, NY 10027, USA*

Phone: 001 212 854 2841

Fax: 001 212 854 3304

Email: da2203@columbia.edu

Website: http://www.me.columbia.edu/lmtp

**Abstract**

We demonstrate and explain a simple and efficient way to remove gas bubbles from liquid-filled microchannels, by integrating a hydrophobic porous membrane on top of the microchannel. A prototype chip is manufactured in hard, transparent polymer with the ability to completely filter gas plugs out of a segmented flow at rates up to 7.4 μL/s per mm$^2$ of membrane area. The device involves a bubble generation section and a gas removal section. In the bubble generation section, a T-junction is used to generate a train of gas plugs into a water stream. These gas plugs are then transported towards the gas removal section, where they slide along a hydrophobic membrane until complete removal. The system has been successfully modeled and four necessary operating criteria have been determined to achieve a complete separation of the gas from the liquid. The first criterion is that the bubble length needs to be larger than the channel diameter. The second criterion is that the gas plug should stay on the membrane for a time sufficient to transport all the gas through the membrane. The third criterion is that the gas plug travel speed should be lower than a critical value: otherwise a stable liquid film between the bubble and the membrane prevents mass transfer. The fourth criterion is that the pressure difference across the membrane should not be larger than the Laplace pressure to prevent water from leaking through the membrane.

## *1. Introduction*

Bubbles can be generated in microfluidic systems continuously by flow-focusing (Garstecki et al. 2004; Gordillo et al. 2004; Cubaud et al. 2005; Sevilla et al. 2005; Hettiarachchi et al. 2007; Hashimoto et al. 2008; Herrada and Ganan-Calvo 2009) and T-junction configurations (Laari et al. 1997; Gunther et al. 2005) or on-demand by thermal heating (Prakash and Gershenfeld 2007) and piezo actuation (Xu and Attinger 2008). Sometimes unwanted gas pockets can form accidentally due to priming or cavitation. These bubbles are sometimes useful, e.g. enhancing heat and mass transfer (Gunther et al. 2004; Kreutzer et al. 2005; Betz and Attinger 2009), creating microstreaming (Kao et al. 2007), providing a platform for biochemical synthesis (Choi and Montemagno 2006), enhancing mixing for chemical reaction and cell lysis (Gunther et al. 2004; El-Ali et al. 2005). Most of the times, however, bubbles are associated with disturbances in microfluidic devices. For instance, they can clog channels (Jensen et al. 2004) or reduce the dynamic performance (van Steijn et al. 2007) of the microfluidic device. Furthermore, exhaust gas bubble generation is known deteriorate the performance of microchannel based micro fuel cells (Kamitani et al. 2008; 2009; Paust et al. 2009). Therefore a gas removal process integrated to the chip is of high interest in microfluidics. Various methods have been explored for trapping and removing bubbles from a microchannel, such as dynamic bubble traps (Schonburg et al. 2001), and diffusion/capillarity based devices (Gunther et al. 2005; Meng et al. 2006; Skelley and Voldman 2008; Sung and Shuler 2009; Zhu 2009). Dynamic bubble traps are often used in extracorporeal blood flow circuits: they use 3D spiral channels to accelerate the flow radially and focus the bubbles towards one location, where extraction proceeds (Schonburg et al. 2001). However, a significant amount of liquid might be extracted together with the gas. Diffusion-based bubble removal has been successfully shown using a gas-permeable membrane, such as a thin PDMS layer as in (Skelley and Voldman 2008; Sung and Shuler 2009). However, the reported gas removal rates are relatively low, typically $1 \times 10^{-4}$ µL/s per mm$^2$ (Sung and Shuler 2009). Thus, for reported practical applications, at least several minutes (Skelley and Voldman 2008; Sung and Shuler 2009) of extraction time are needed. Alternatively, a porous membrane can be used to separate immiscible fluids: Kralj and co-workers (Kralj et al. 2007) achieved complete separation of organic-aqueous and fluorous-aqueous liquid/liquid systems in a microfluidic device, and provided two design criteria for successful separation. Gas/liquid separation using porous membrane has also been reported (Gunther et al. 2005; Meng et al. 2006; Kamitani et al. 2009; Paust et al. 2009; Zhu 2009). For example, Zhu (Zhu 2009) demonstrated that hydrophobic and hydrophilic membranes can be used together in the end of a microchannel to achieve a complete gas/liquid separation by letting gas and liquid flow through hydrophobic and hydrophilic membranes respectively, but the study did not mention the gas removal rate. He also achieved incomplete separation by using a hydrophilic membrane in a channel flown with a gas/water mixture. Similarly, Kamitani (Kamitani et al. 2009) used a hydrophilic porous membrane to enhance liquid filling through the membrane and gas detachment from the membrane in a direct methanol fuel cell. With the help of hydrophobic venting, Meng and Kim realized a micropump by directionally controlled bubble growth (Meng and Kim 2008). In terms of modeling, several studies have performed calculations of leakage pressure (Kralj et al. 2007; Meng et al. 2007; Zhu 2009). Also Meng and coauthors built a model for the bubble venting rate through the membrane (Meng et al. 2007). However, to the best of our knowledge, there is currently no complete set of physics-based design rules to describe gas removal using hydrophobic membranes in microfluidic devices. In this experimental study, we integrate a hydrophobic membrane into a microfluidic chip and successfully separate gas plugs from a segmented flow. We also investigate the theoretical conditions for bubble extraction, and provide four criteria to be satisfied in order to achieve completed separation of the gas from the liquid.

## *2. Material, fabrication and assembly*

Figure 1 shows the assembly of the gas removal device used in our experiment. The microchannels are 500 µm wide and 500 µm deep. They are milled out from a PMMA (Polymethylmethacrylate) using a Minitech CNC milling machine, with less than 500 nm surface roughness. The channels are then sealed with PMMA, the 200 μm thick hydrophobic acrylic copolymer membrane (Pall Corporation), and 70 μm thick double-sided tape (Adhesives Research, Inc), as shown in Figure 1.

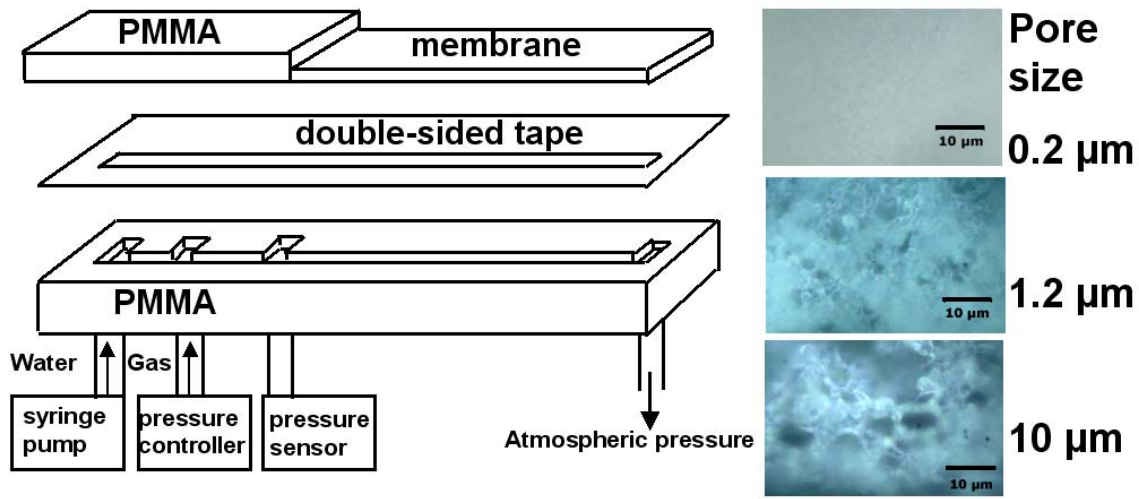

Figure 1: (Left) Assembly of bubble removal chip. A 500 μm wide slit is cut through the tape, and aligned on top of the main channel. Bubbles are generated at a T-junction, where water is pushed by syringe pumps (KDS 210) and the gas pressure is controlled by a pressure controller (Druck DPI 530, 2 bar gauge, precision ±1% FS). The generated bubbles are then transported to the porous membrane, where extraction takes place. (Right) Micrographs of porous hydrophobic membranes with different pore sizes.

The three tested porous hydrophobic membranes (Pall Corporation) are made of acrylic copolymer and had three respective typical pore sizes, 0.2, 1.2 and 10 μm (Figure 1). A 500 μm wide slit is cut through the tape, and aligned on top of the main channel. Therefore, the bubble generation section has all four walls made of PMMA while the gas removing section has a channel made of three PMMA walls and one membrane wall, if we neglect the presence of the tape. In the bubble generation section, gas plugs are generated at a T-junction, where water is pushed by syringe pumps (KDS 210) and the gas pressure is controlled by a pressure controller (Druck DPI 530, 2 bar gauge, precision ±1% FS). The generated bubbles are then transported to the bubble removing section, where extraction takes place. A piezoresistive pressure sensor (Honeywell ASCX15DN, 103.4 kPa differential, repeatability ±0.2% FS) is used to monitor the pressure difference between the atmosphere and the fluid upstream of the hydrophobic membrane.

## *3. Experimental results*

To generate gas plugs at the T-junction with different speed, backpressure in the gas and water flow rate were varied. We found that bubbles smaller than the channel diameter could not be removed. Therefore, we focused on generating gas plugs longer than the channel height, to ensure that they are constrained by the microchannel. The measured void fraction ranged from 0.25 to 0.78. The picture sequence in Figure 2 gives a close look at the bubble dynamics during the removal process. We observe that the receding contact angle at the bubble front increases during the vanishing period, as shown in frames 0 to 4.4ms. Also the vanishing bubble first reduces its length, for example the bubble at 2.8 ms is about half of the original length. Then, after 3.2 ms, the height of the bubble starts to decrease. While the contact area between gas and membrane is also decreasing, the remaining part of the bubble seen from the side assumes a sharp triangular shape, before it fully disappears (see e.g. frames at 0.8 and 1.2ms). We found out that this interesting curvature change only occurs when the Weber number is greater than unity, as shown in Figure 3. The reason for this is probably that due to the competition of inertial forces and surface tension forces, in the sense that flow over the bubble creates a pressure difference between the back and the front of the bubble large enough to be induce a change of curvature.

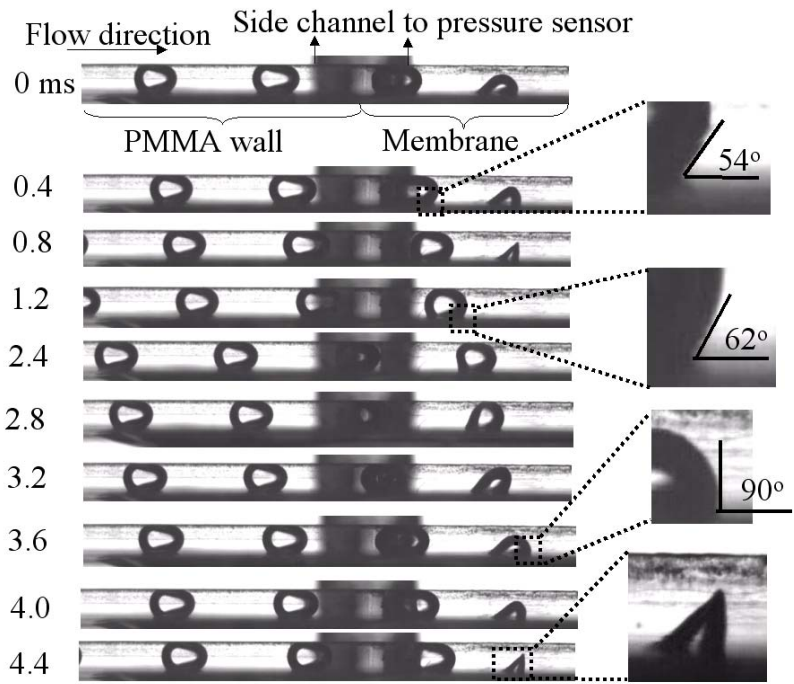

Figure 2: Picture sequence of a typical bubble removal process, using a membrane with 1.2 μm pores. Bubbles are traveling at a speed of 0.62 m/s and are completely removed from the channel through the membrane.

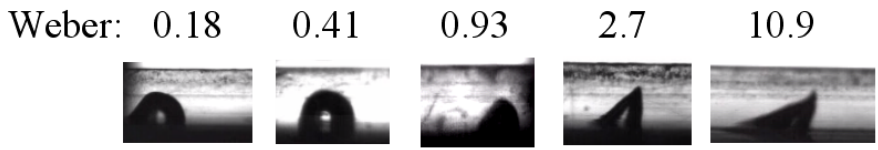

Figure 3: Shapes of vanishing bubbles at different Weber numbers. The strange shape at large Weber number is due to the pressure drop along the bubble length.

We also measured the flow rates and gage pressure needed for water to leak through the porous membrane, by flowing pure water in the channels, while the gas inlet was closed. While for 0.2 μm pore size membrane, our syringe pump fails at 46 mL/m and a gage pressure of 81 kPa, leakage of water through the membrane was found to occur at 40 and 21 mL/m of water flow rate for 1.2 and 10μm pore sizes respectively, and the respective gage pressures, measured at the location upstream from the membrane, were 41 and 20 kPa. For all experiments described below, the gage pressure was kept lower than these critical pressures to prevent water leakage. We found that for complete gas extraction cases, a maximum extraction rate of about 7.4 µL/s/mm$^2$ is achieved with a 10 µm pore-sized membrane of 60 mm$^2$ exposed area. Such extraction rate is four orders of magnitude higher than previously reported using a PDMS membrane (Sung and Shuler 2009). This enhancement is probably due to the different gas transport mechanisms. Across PDMS, gas transport is due to solution and diffusion, and the steady-state gas mass flux $N$ (kg/m$^2$/s) obeys the equation (Merkel et al. 2000):

$$N = \frac{P \Delta p}{h} \qquad (1)$$

where $h$ is the membrane thickness, $\Delta p$ is the pressure difference across the membrane and $P$ is the gas permeability, which is 1.34x10$^{-16}$ kmol/(Pa s m$^2$) for Nitrogen. On the other hand, gas transport through a porous membrane, is due to the viscous flow in the parallel pores, and the steady-state gas volume flux $q$ (m$^3$/m$^2$/s) can be estimated from Darcy's law (Barrer 1967):

$$q = \frac{\kappa}{\mu} \frac{\Delta p}{h} \qquad (2)$$

where $\mu$ is the gas viscosity, and $\kappa$ is permeability of the membrane, which has been obtained experimentally as follows. By varying the airflow $q$ through the membrane, the pressure drop across the membrane is recorded and plotted in Figure 4. According to Figure 4, $\kappa$ is calculated to be 7.8x10$^{-15}$, 2.9x10$^{-13}$ and 1.3x10$^{-12}$ m$^2$ for membranes with 0.2, 1.2 and 10 µm pores respectively. Calculations in Table 1 reveals that, under the same pressure across the membrane, the mass/volume flux in porous membrane can be four orders of magnitude higher than in a PDMS membrane with the same thickness.

|  | Pore size $D$ (μm) | Permeability $\kappa$ (m$^2$) | Thickness $h$ (μm) | Pressure drop across membrane $\Delta p$ (kPa) | Mass flux $N$ (kg/m$^2$/s) | Volume flux $q$ (m$^3$/m$^2$/s) | Achieved gas removal rate (μL/s/mm$^2$) |
|---|---|---|---|---|---|---|---|
| Porous membrane | 0.2 | 7.8x10$^{-15}$ | 200 | 10 | 1.6x10$^{-5}$ | 2.1x10$^{-2}$ | 6.3x10$^{-1}$ |
|  | 1.2 | 2.9x10$^{-13}$ |  |  | 6.1x10$^{-4}$ | 8.1x10$^{-1}$ | 5.6 |
|  | 10 | 1.3x10$^{-12}$ |  |  | 2.7x10$^{-3}$ | 3.6 | 7.4 |
| PDMS membrane | N/A |  | 200 | 10 | 1.9x10$^{-7}$ | 2.5x10$^{-4}$ | 5x10$^{-4}$ |

**Table 1: theoretical mass flow rate across a porous membrane and a PDMS membrane.**

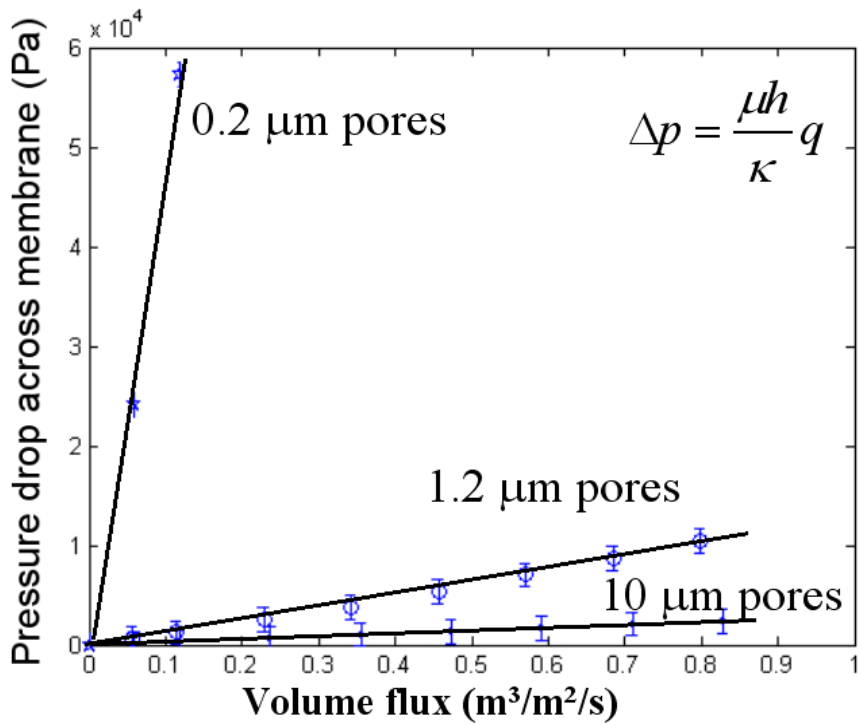

Figure 4: To determine the permeability $\kappa$ of the membranes, the pressure drop across the membrane is measured and plotted as a function of the volume flux of the airflow through the membrane.

### *4. Discussions*

Two outcomes are unsatisfactory for a gas removal device: membrane leakage and incomplete extraction. During incomplete extraction, the outflow is not pure water but an air-liquid mixture. During membrane leakage, water and gas go through the membrane.

Our analysis below shows that four criteria, as listed in Table 2, need to be simultaneously satisfied to guarantee complete gas extraction without membrane leakage.

| Criterion | Equation | Physical meaning |
|---|---|---|
| 1 | $L_{bubble} > H$ | Bubble length $L_{bubble}$ needs to be larger than the channel height $H$. |
| 2 | $L/v > \tau = H \dfrac{\mu}{\kappa} \dfrac{h}{p_b - \dfrac{4\gamma}{H} - p_0} \ln(\dfrac{l_0}{l_1})$ | Bubble traveling time on a membrane that has a length of $L$ should be sufficient to transport all the gas through the membrane. |
| 3 | $v < v_{c\_mmebrane} = \dfrac{1}{9\sqrt{3}} \dfrac{\gamma/\mu}{a} \theta_{E\_membrane}^3$ | Bubble speed $v$ should be lower than a critical value: otherwise a stable liquid film between the bubble and the membrane prevents mass transfer. |
| 4 | $\Delta p < \Delta p_{LP} = -\dfrac{4\gamma \cos\theta}{d}$ | The pressure difference across the membrane $\Delta p$ should be smaller than the Laplace pressure $\Delta p_{LP}$ to prevent water leakage. |

Table 2: These four criteria need to be simultaneously satisfied to successfully remove gas bubbles from microfluidic channels

### 4.1 Criterion 1, 2 and 3: complete gas extraction

As mentioned above, the geometry of our bubble trap requires a bubble length larger than the channel height for putting bubble and membrane in contact, thus allowing for degassing. This is the first necessary criterion for complete gas extraction, criterion 1 in Table 2. However, in practice, gas bubbles smaller than the channel could also be removed with the help of a channel contraction or an obstacle that causes the flowing bubble to contact the channel. A second criterion can be formulated by considering the time needed to fully extract the gas bubble. We can equal the shrinking rate of the bubble $dV/dt$ to the gas flow rate $Q$ through the membrane, which can be estimated by Darcy's law (equation (2)):

$$-\frac{dV}{dt} = Q = \frac{\kappa}{\mu} \frac{p_b - p_0}{h} \frac{V}{H}, \qquad (3)$$

where $V/H$ gives the contact area between the bubble and the membrane, $\kappa$ the permeability given in Table 1 $p_b$ is the pressure in the bubble and $p_0$ is the atmospheric pressure. In the experiment, $p_b$ is estimated by $p_l + \frac{1}{2}(\frac{2\gamma}{r_1} + \frac{2\gamma}{r_2})$, where $p_l$ is the liquid pressure measured by pressure sensor and $r_1$, $r_2$ are the radii of curvature measured at the bubble head and tail respectively. Assuming that the bubble shrinks by reducing its length keeping its pressure and height constant, we can integrate to determine the extraction time $\tau$:

$$\tau = H \frac{\mu}{\kappa} \frac{h}{p_b - p_0} \ln(\frac{l_0}{l_1}) \qquad (4)$$

where $l_0$ and $l_1$ are the initial bubble length and final bubble length respectively. This integral does not converge to a finite time, however the analysis corresponds well to most of the extraction process, so that a reasonable estimate of the bubble extraction time is obtained by assuming a small value $l_1$ of 1% of the channel height. Figure 5 plots the comparison between experiments and the theory, which shows good agreements for the membrane with 0.2 μm pores. Therefore, to ensure complete extraction, the bubble should move along the membrane for a time no shorter than. However, Equation 4 is not very convenient to be used for design purposes, because it requires knowledge of the bubble curvature and of the pressure, rather than just an estimate of the pressure in the liquid $p_b$. We observe that the Laplace contribution is bounded to a range, i.e. $-\frac{4\gamma}{H}$ to $\frac{4\gamma}{H}$, with a worst-case scenario happening when $p_b = p_l - \frac{4\gamma}{H}$, because a lower pressure in the bubble always slows down the bubble removal. Therefore, for design purposes, we use can estimate $p_b$ as $p_l - \frac{4\gamma}{H}$, as listed in Table 2.

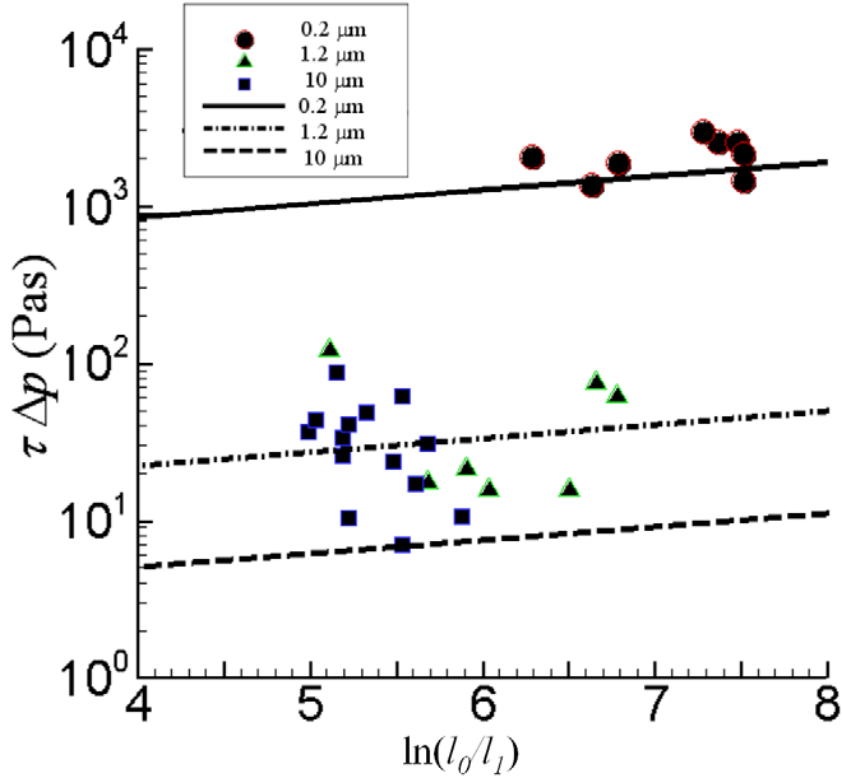

Figure 5: Comparison between experiments and the theory given by Equation 4 for criteria 2. The x-axis plots the term $\ln(l_0/l_1)$ and the y-axis plots the product of the extraction time and the pressure. We can see a good agreement for the membrane with 0.2 μm pores. However, for the membranes with 1.2 and 10 μm pores, the extraction time are larger than theoretical prediction, which can be explained by the liquid film formed between the wall and the gas plug, as analyzed in Criteria 3.

For the membranes with 1.2 and 10 μm pores, the theoretical $\tau$ can be on the order of milliseconds, thus this criterion suggests that gas bubbles can be removed from very fast flows, at speed on $10^4$ m/s, much faster than in the experiments described here. This situation is unrealistic because of the coating liquid films surrounding bubbles traveling in channels at non-negligible capillary numbers. In this case, a third criterion has to be formulated for complete gas removal to account for the liquid film between the wall and the gas plug, which might delay, as seen in Figure 5, or compromise gas extraction. A static gas/water interface would contact the wall with a contact angle $\theta_E$ and form a clear triple line. However, an interface moving along the wall will exhibit a dynamic contact angle $\theta_D$, which decreases for increasing bubble velocities. There is therefore a critical velocity, where the wetting angle approaches zero, and above which a film appears between the plug and the wall because the triple line cannot find a stable position anymore. The critical velocity $v_c$, can be estimated by (de Gennes et al. 2003):

$$v_c = \frac{1}{9\sqrt{3}} \frac{\gamma/\mu}{a} \theta_E^3 \qquad (5)$$

where $a = 20$ is a dimensionless coefficient that only weakly depends on $v$. For an air-water system in respective contact with PMMA and membrane surfaces, $v_c$ is calculated to be 0.38 and 2.3 m/s, respectively, using contact angles measured in the experiments (68° for PMMA, 124° for the porous membrane). Once the film is formed, the thickness $e$ of the film can be calculated as (de Gennes et al. 2003):

$$e = \frac{D}{2} Ca^{2/3} \qquad (6)$$

where $D$ is the channel diameter and $Ca$ is the capillary number. Assuming $D$ as the hydraulic diameter of our channel, Figure 6 (second y-axis) shows the theoretical film thickness $e$ as a function of bubble speed $v$ on both PMMA and membrane surfaces. In our experiment, bubbles travel along the PMMA wall and then on the membrane so that the corresponding film situations occur as in Table 3:

| Bubble speed $v$ | Film thickness on PMMA | Film thickness on membrane | Membrane length needed for complete gas removal |
|---|---|---|---|
| $v < v_{c\_PMMA}$ | no film | no film | ~ 0 |
| $v_{c\_PMMA} < v < v_{c\_membrane}$ | $e = \frac{D}{2} Ca^{2/3}$ | Decreasing until rupture | Finite |

| $v > v_{c\_membrane}$ | $e = \dfrac{D}{2} Ca^{2/3}$ | $e = \dfrac{D}{2} Ca^{2/3}$ | Infinite |

Table 3: film situations for a gas plug traveling along the PMMA wall first and then on the membrane

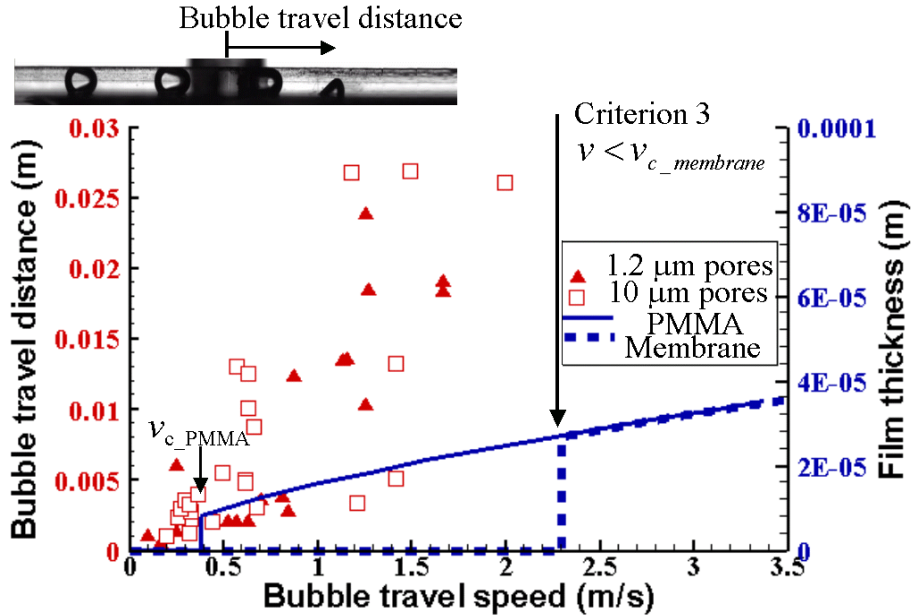

Figure 6: Experimental maximum bubble travel distance and theoretical film thickness as functions of bubble travel speed. For the bubbles that are slower than $v_{c\_membrane}$ can be completely extracted at certain locations in the channel, a situation that was not achieved for bubbles that are faster than $v_{c\_membrane}$. Note that, the bubble travel distance generally increases with the bubble travel speed, however the data points look scattered. This may be due to the nonuniformity of the pore sizes and nonhomogeneous distribution of the pores on the membrane surface, as pictured in Figure 1.

According to Table 3, if the bubble speed $v$ is greater than $v_{c\_membrane}$, a stable film between the bubble and the membrane will prevent gas removal. On the other hand, if the bubble speed $v$ is smaller than $v_{c\_membrane}$, the film might become unstable on top of the membrane and rupture so that gas can be removed, provided the membrane is long enough. In the experiments reported using first y-axis in Figure 6, we see that gas plugs that are slower than $v_{c\_membrane}$ can be completely extracted at certain locations in the channel, a situation that was not achieved for gas plugs that are faster than $v_{c\_membrane}$.

Note that, the bubble travel distance generally increases with the bubble velocity, however the data points look scattered. This may be due to the nonuniformity of the pore sizes and nonhomogeneous distribution of the pores on the membrane surface, as pictured in Figure 1. Though the definition of surface roughness of porous media is not very straightforward (Hermann et al. 1992), we believe that the rough surface topology of the membrane can have three major effects: 1) surface roughness tends to increase the macroscopic contact angle, or apparent contact angle (de Gennes et al. 2003). Therefore, we measured the macroscopic contact angle from a sessile drop on the membrane and used this measured value in Equation 5; 2) the air-filled pores under the membrane surface tend to promote film rupture on the membrane surface (Slavchov et al. 2005); 3) surface roughness can cause contact line pinning and depinning during advancing and receding (Duursma et al. 2009). While these three effects add complexity to our physical model, the approximation in Criterion 3 is meaningful, and sufficient to provide guidance for design purposes.

## 4.2 Criterion 4: membrane leakage

A porous hydrophobic membrane will prevent the water-air meniscus to go through the pores because of interfacial tension, a situation analyzed in (Kralj et al. 2007) for a liquid/liquid system. Using the same principle, we formulate a criterion necessary to prevent water from leaking through a porous hydrophobic membrane.

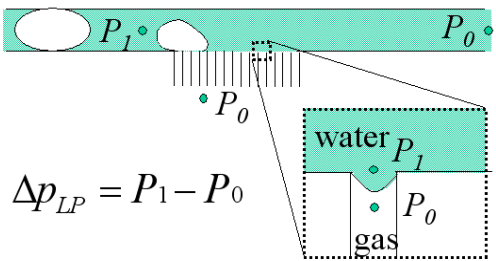

Figure 7: Bubble extraction working principle: an air-liquid meniscus is pinned at the entrance of the pore and the surface tension holds the pressure difference across the meniscus and prevents water from leaking through the pore.

Figure 7a shows the air-liquid meniscus is pinned at the entrance of the pore and the surface tension holds the pressure difference across the meniscus and prevents water from leaking through the pore. However, with an increasing pressure difference, the angle between the meniscus and the inner wall of the pores will reach a maximum value of the equilibrium wetting angle $\theta$. In another word, a meniscus can hold a pressure difference up to a maximum value of

$$\Delta p_{LP} = -4\gamma \cos\theta / d , \qquad (7)$$

where $\gamma$ is the surface tension between gas and water, $\theta$ is the contact angle and $d$ is the pore size (de Gennes et al. 2003). As long as the pressure difference $\Delta p$ across the membrane is smaller than $\Delta p_{LP}$, there will be no water leaking through the membrane, which gives our fourth criterion as listed in Table 2. Using the pore size given by the manufacturer, the Laplace pressures $\Delta p_{LP}$ are calculated as 804, 134 and 16 kPa for 0.2, 1.2 and 10 µm membrane respectively, while in the experiments, water starts to leak at >81(where our syringe pump fails), about 41 and 20 kPa respectively. These values are reasonable considering uncertainties on the pore shape and size (see Figure 1) or on the wetting angle (de Gennes et al. 2003).

## *5. Conclusion*

A microfluidic device has been manufactured to separate gas from water in a segmentation flow. Four necessary operating criteria have been determined experimentally and explained theoretically to achieve a complete separation of the gas from the liquid: 1) the bubble length should be larger than the channel diameter; 2) the gas plug should stay on the membrane for a time no shorter than a critical value; 3) the bubble speed should be lower than a critical value; 4) the pressure difference across the membrane should be lower than a critical value. The corresponding equations for these criteria are listed in Table 2. To further investigate the bubble dynamics and the separation physics, we plan in future work to use computational fluid dynamics to simulate the two-phase flow along and across the porous membrane, in a complex geometry.

**Acknowledgement:**

This study is supported by NSF CAREER award 0449269. We also thank Pall Corporation for graciously providing the membranes.